\documentclass[twocolumn, secnumarabic, amssymb, aps, showpacs, prd]{revtex4-2}

\usepackage{graphics}
\usepackage[dvips]{graphicx}
\usepackage{dcolumn}
\usepackage{bm}
\usepackage{xcolor}
\usepackage{microtype}
\usepackage{amsmath}
\usepackage{setspace}
\usepackage{mathrsfs}
\usepackage{geometry}
\usepackage{listings}
\usepackage{xcolor}
\usepackage{enumitem}
\usepackage{hyperref}

\begin{document}
\title{\textbf{On Energy References when a Magnetic Field is Applied.}\\
 A Practical Guide to Reference Energies with and without a Magnetic Field}
\author{Pedro Pereyra}
\affiliation{Departamento de Ciencias B\'asicas, UAM-Azcapotzalco, M\'{e}xico D.F., C. P. 02200, M\'exico}
\date{\today}

\begin{abstract}
We present a simple analysis of the energy references and of the magnetic-field-dependent and system-dependent features of the Landau energy levels. The energy reference used in describing Landau quantization is usually taken as given, with the Fermi levels of the electron system and the coupled reservoir aligned in thermal equilibrium. Nevertheless, the dependence of the Landau-level spectrum on magnetic field can make the relation between these energy references less transparent.
In this work, we show that when a magnetic field is applied and the confined Landau levels $E_n$ are established, both the Fermi level $E_F$ and the bottom energy $E_0$ remain fixed when the magnetic field varies. We show that the bottom energy $E_0$ is independent of $B$, and determined by a system-dependent field $B_1$, defined by the condition that the first Landau level equals the Fermi energy. In contrast, the minimum of the parabolic well depends explicitly on the magnetic field. The resulting picture provides a simple way of displaying the relative motion of the Landau levels and the fixed Fermi level without introducing a magnetic-field-dependent energy zero.

\end{abstract}

\maketitle

\section{Introduction}

When a charged particle is in a two-dimensional system and a perpendicular magnetic field is applied, it is well known that the energy eigenvalues of the Landau model constitute an equally spaced spectrum of energy levels \cite{LandauLifshitz}. It is also well known that, as the magnetic field increases, the spacing between consecutive levels increases. When dealing with these energies, it is natural to ask: which is the appropriate energy reference? Is it at the Fermi level? Is it at the minimum of the confining parabolic potential? Is it at the bottom of the Landau spectrum? More generally, where are the Landau levels located with respect to the Fermi energy? These questions may appear elementary, but their answer is not apparent from the Landau spectrum itself, since the spectrum is obtained from the model before its energy reference is related to that of the electron system and its reservoir. The purpose of this brief analysis is to answer these questions.

The Landau spectrum depends on the magnetic field through the cyclotron frequency $\omega_c$ and is commonly described as
$$
E_n=E_0+n\hbar\omega_c,
$$
where $E_0$ denotes the bottom energy of the Landau spectrum. In the usual harmonic-oscillator description, this energy is related to the minimum $E_{\rm min}$ of the corresponding parabolic confining potential by
$$
E_0=E_{\rm min}+\frac{1}{2}\hbar\omega_c.
$$
At the same time, when the electron system is coupled to an external reservoir, before a magnetic field is applied, the Fermi levels of the two systems are aligned in thermal equilibrium. The reservoir therefore provides the relevant Fermi-energy reference for the electron system \cite{DAgostaRaimondiVignale}. The question is where the Landau spectrum is energetically located when the magnetic field is applied. Where is the bottom energy $E_0$? Is it at, above, or below the Fermi energy? Does the distance $E_F-E_0$, if any, depend on the magnetic field? In other words, how does the magnetic parabolic confinement relate to the fixed Fermi level? Does the magnetic field move the minimum of the confining potential, and how is this related to the zero-point energy?
The relation between these energy references deserves some clarification when the magnetic field is varied. The distinction is particularly relevant in teaching the quantum Hall effect, where the Landau spectrum, the Fermi level, and the magnetic confinement potential are often introduced using different energy references without explicitly discussing their relation.

In this work, we review the well-known relations in the Landau model and show that, using a characteristic magnetic field $B_1$, defined by the coincidence of the first Landau level with the Fermi level, the bottom energy $E_0$ of the Landau spectrum is determined as a system-dependent quantity. We also show that, when the magnetic field is subsequently varied, $E_0$ remains fixed for that system, whereas the Landau-level separations and the zero-point cyclotron energy vary with the magnetic field.

We further consider the effective parabolic confinement associated with the magnetic field. In this representation, the minimum of the parabolic well, $E_{\rm min}$, is distinct from the fixed energy $E_0$ and acquires an explicit dependence on the magnetic field through the zero-point cyclotron energy. This distinction provides a simple physical picture in which the Landau levels are positioned relative to a fixed Fermi level while the energy reference itself is not changed.

The purpose of this analysis is therefore not to modify the standard Landau quantization, but to make explicit the different roles played by the fixed system-dependent energy $E_0$, the reservoir-fixed Fermi level $E_F$, and the magnetic-field-dependent minimum $E_{\rm min}$ of the parabolic confining potential. This separation leads naturally to a consistent representation of the energy spectrum and of the magnetic parabolic well using the same energy reference.

\begin{figure*}[hbt]
\begin{center}
\includegraphics[width=16.5cm]{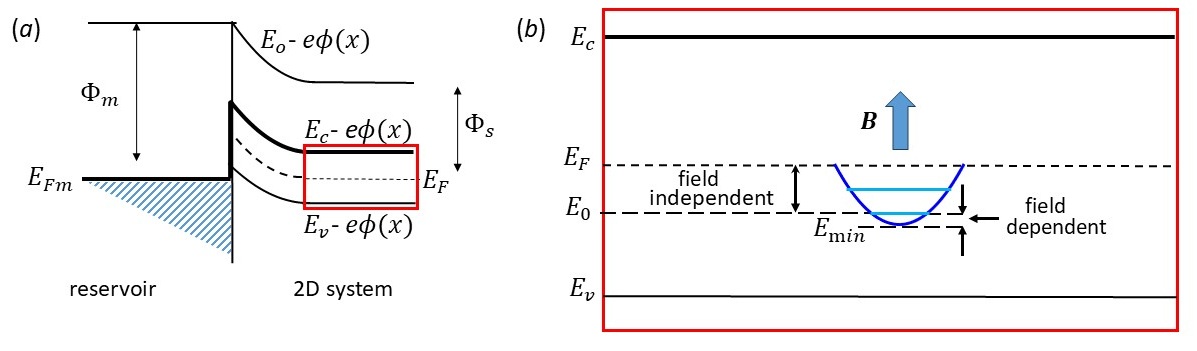}
\caption{Schematic representation of reference energies. In a) the reservoir and electronic  Fermi energy alignment. In b) the red square in a) enlarged to show the  local representation of the parabolic magnetic confinement, Landau energy levels, the bottom energy $E_0$ and the minimum energy $E_{\rm min}$ of the parabolic well. We also indicate the energy spacings that are fixed (independent of the magnetic field) and the field dependent spacing between the minimum and the bottom energy. }
\label{energyreferences}
\end{center}
\end{figure*}
\section{Landau spectrum bottom $E_0$ and the parabolic well minimum $E_{\rm min}$}

Before the magnetic field is applied, the graphene sample is in contact with a reservoir. Consequently, the Fermi level of the sample aligns with the Fermi level of the reservoir, and we denote their common value by $E_F$. The Fermi energy is therefore fixed by the reservoir and does not move when the magnetic field is subsequently varied.

For a graphene sample whose Fermi level lies in the middle of the gap,
one may write
\[
E_F=E_v+\frac{E_g}{2},
\]
where $E_v$ is the valence-band energy and $E_g$ is the energy gap. More
generally, the midpoint of the gap is
\[
E_F=\frac{E_c+E_v}{2},
\qquad
E_g=E_c-E_v.
\]

When a magnetic field is applied, it does not move the Fermi level.
Instead, it gives rise to an effective confining potential, which may be viewed
as a parabolic quantum well whose minimum $E_{\rm min}$ is determined by the magnetic
field. Within this quantum well, a quantized Landau spectrum $E_n$ is established,
whose bottom $E_0$ is determined by the system-dependent energy associated with the QHE.
Let us now determine these energies.

\subsection{The bottom energy $E_0$ of the spectrum}

We will see here that, while the presence of the magnetic field does give rise to a shallow parabolic potential where the Gaussian-Hermite wave packets $\eta_n$, corresponding to the Landau spectrum $E_n$, are confined (see figure \ref{energyreferences}), it does not determine the bottom energy $E_0$ of the spectrum, nor does it determine the Fermi level $E_F$.

Let $B_1$ denote the magnetic field for which the first Landau level
aligns with the Fermi level:
\[
E_1(B_1)=E_F.
\]
The field $B_1$ is not an arbitrary theoretical parameter. It is a
system-dependent quantity fixed independently through the QHE
experiment. Once the physical system is specified and $B_1$ is
determined experimentally, the corresponding bottom energy $E_0$ is
also fixed, and we shall show it now.

In the present convention, the Landau levels are written as
\[
E_n=E_0+n\hbar\omega_c,
\qquad
\omega_c=\frac{qB}{m}.
\]
At the field $B_1$, the condition
\[
E_1(B_1)=E_F
\]
gives
\[
E_F=E_0+\hbar\frac{qB_1}{m},
\]
and hence
\[
\boxed{
E_0(B_1)=E_F-\hbar\frac{qB_1}{m}.
}
\]
The field $B_1$ therefore fixes the bottom energy of the Landau
spectrum for the given system.

Now consider a different magnetic field
\[
B_n=\frac{B_1}{n}.
\]
At this field, the $n$-th Landau level aligns with the same fixed Fermi
level:
\[
E_n(B_n)=E_F.
\]
Using the Landau-level expression,
\[
E_n(B_n)
=
E_0+
n\hbar\frac{qB_n}{m}.
\]
Since
\[
B_n=\frac{B_1}{n},
\]
we obtain
\[
E_n(B_n)
=
E_0+\hbar\frac{qB_1}{m}
=
E_F.
\]
Thus,
\[
\boxed{
E_0(B_n)=E_0(B_1)
}
\]
for the sequence of fields $B_n=B_1/n$ at which successive Landau
levels align with $E_F$. For example, if
\[
B_2=\frac{1}{2}B_1,
\]
then
\[
E_2(B_2)
=
E_0+
2\hbar\frac{qB_2}{m}
=
E_0+\hbar\frac{qB_1}{m}
=
E_F.
\]
Therefore,
\[
E_0
=
E_F-2\hbar\frac{qB_2}{m}
=
E_F-\hbar\frac{qB_1}{m}
=
E_0(B_1).
\]
This shows that the bottom energy of a given physical system does not
move when one passes from one QHE crossing field $B_n$ to another.
Different values of $B_n$ correspond instead to different Landau
levels reaching the same fixed Fermi level. Different physical systems
may have different values of $B_1$ and therefore different values of
$E_0$.

\subsection{The zero-point energy and the minimum $E_{\rm min}$ of the confining potential}

There is an additional feature of the confining parabolic potential due to the presence of the magnetic field: the zero-point energy. The Landau-level energy contains the term
$$
\frac{1}{2}\hbar\omega_c
=
\frac{1}{2}\hbar\frac{qB}{m}.
$$
This term is the zero-point energy, independently of any particular
definition of the total energy $E_T$. It is an intrinsic consequence of
the quantization of the cyclotron motion. It depends on the magnetic
field $B$ and therefore increases as $B$ increases.

At the special field $B_n=B_1/n$, the $n$-th Landau level can therefore
be written as
$$
E_n(B_n)
=
\frac{1}{2}\hbar\frac{qB_n}{m}
+
n\hbar\frac{qB_n}{m}.
$$

Since
$$
n\hbar\frac{qB_n}{m}
=
\hbar\frac{qB_1}{m}
=
E_F-E_0(B_1),
$$
we obtain
$$
\boxed{
E_n(B_n)
=
\frac{1}{2}\hbar\frac{qB_n}{m}
+
\left[E_F-E_0(B_1)\right].
}
$$

Notice the different roles of the quantities in this expression. The
first term,
$$
\frac{1}{2}\hbar\frac{qB_n}{m},
$$
depends on the magnetic field and is the zero-point energy. The second
quantity,
$$
E_F-E_0(B_1),
$$
is fixed for the given physical system because $E_F$ is fixed by the
reservoir and $E_0(B_1)$ is fixed by the experimentally determined
system parameter $B_1$.

Thus, as the magnetic field changes, the zero-point contribution changes
and grows with $B$, while the Fermi level and the system-dependent
bottom energy remain fixed.

The minimum of the magnetic quantum well can consequently be
represented in terms of the bottom energy as
$$
\boxed{
E_{\rm min}(B)
= E_0(B_1)
-
\frac{1}{2}\hbar\frac{qB}{m}.
}
$$

Here the first term is fixed for the given system through $B_1$, whereas
the second term is the magnetic-field-dependent zero-point contribution.
Thus, as $B$ increases, the minimum $E_{\rm min}$ moves downward relative
to the fixed $E_0$, while the distance between $E_{\rm min}$ and $E_0$
increases with $B$.

The important distinction is therefore that the magnetic field does
not move the reservoir Fermi level:
$$
E_F=\mathrm{constant}.
$$
It changes the Landau-level spacing and the zero-point contribution,
while the experimentally determined $B_1$ fixes the characteristic
bottom energy $E_0(B_1)$ of the given system.

As the magnetic field becomes smaller, the Landau-level spacing
$$
\hbar\omega_c=\hbar\frac{qB}{m}
$$
becomes smaller. Consequently, more Landau levels can lie between
$E_0(B_1)$ and the fixed $E_F$. At the field $B_1$, the first excited
level reaches $E_F$:
$$
E_1(B_1)=E_F.
$$

At
$$
B_2=\frac{B_1}{2},
$$
the second level reaches $E_F$:
$$
E_2(B_2)=E_F.
$$
More generally,
$$
B_n=\frac{B_1}{n},
\qquad
E_n(B_n)=E_F.
$$
Thus, for smaller magnetic fields, the smaller Landau-level spacing
allows an increasing number of quantized levels to occur between the
fixed bottom energy and the fixed Fermi level.

\subsection{The total energy in the presence of a magnetic field}

Finally, one may define the total energy quantity
$$
\boxed{E_T=E_F+\frac{1}{2}\hbar\omega_c.}
$$
This definition is independent of the identification of
$\frac{1}{2}\hbar\omega_c$ as the zero-point energy. The zero-point term
is already present in the Landau quantization itself. In the present
interpretation, $E_T$ is a $B$-dependent physical energy constructed
from the fixed reservoir Fermi energy and the zero-point cyclotron
energy; it is not another energy zero.

The essential picture is therefore the following. A single energy zero
is fixed independently of $B$. The reservoir fixes the sample Fermi
level $E_F$. The QHE experiment determines the system-dependent field
$B_1$, for which $E_1=E_F$, and this fixes the corresponding bottom
energy $E_0(B_1)$. When the field is varied, successive Landau levels
reach the same fixed $E_F$ at fields $B_n=B_1/n$. The zero-point energy
$$
\frac{1}{2}\hbar\frac{qB}{m}
$$
remains explicitly $B$ dependent and increases with the magnetic field,
whereas $E_F$ and $E_0(B_1)$ are fixed for the given system.

\section{Conclusion}

We revisited well-established relations in widely studied two-dimensional
electron systems under a magnetic field and made evident that, among the
three characteristic energy references, the Fermi energy $E_F$, the
Landau spectrum bottom $E_0$, and the parabolic confining potential
minimum $E_{\rm min}$, the first two remain fixed for a given physical
system, whereas the last one depends explicitly on the magnetic field.
In the present convention, $E_{\rm min}$ moves downward as the magnetic
field increases.

In other words, the distance between the Landau spectrum bottom and the
minimum of the parabolic confining potential,
$$
E_0-E_{\rm min}
=
\frac{1}{2}\hbar\frac{qB}{m},
$$
is determined by the magnetic field and increases with it. This analysis
also shows that the bottom energy $E_0$ is a system-dependent quantity
that remains fixed, while the Fermi energy is fixed by the reservoir.
Therefore, $E_0$ and $E_F$ constitute constant reference energy levels
for the given physical system, whereas $E_{\rm min}$ provides the
magnetic-field-dependent reference associated with the parabolic
confining potential.

\section{Acknowledgments}
The author is members of the Sistema Nacional de Investigadoras e Investigadores (SCHTI, Mexico) and acknowledges its economical support.

\end{document}